\documentclass[11pt,a4]{llncs}

\usepackage[T1]{fontenc}
\usepackage[utf8]{inputenc}
\usepackage{amssymb,amsmath}
\usepackage{times}

\usepackage{xspace}

\usepackage{tikz}
\usepackage{pgf}

\usepackage{url}
\usepackage{listings}

\newcommand{\x}{\xspace}
\newcommand{\coq}{\textsc{Coq}\x}
\newcommand{\setQ}{\ensuremath{\mathbb{Q}}\x}

\newcommand{\gp}{\ensuremath{\texttt{gp}\x}}
\newcommand{\bp}{\ensuremath{\texttt{bp}\x}}
\newcommand{\sem}[1]{\ensuremath{[\![#1]\!]\x}}
\renewcommand{\epsilon}{\varepsilon}
\usepackage{lstcoq}
\begin{document}

%% KEYWORDS
% Mobile autonomous robots, gathering, Byzantine tolerance, impossibility
% results, proof certification

%% TOPICS
% Distributed algorithms; correctness and complexity
% Fault tolerance, reliability, availability
% Specification, verification, and testing: tools, methodologies
% Mobile agents and autonomous robots
\thispagestyle{empty}
\begin{tikzpicture}[remember picture, overlay]
\node [xshift=2cm,yshift=0cm]  at (current page.west)
[scale=1.3,very thick, rectangle, draw, fill=gray!50,text width=1cm,text centered]
{$\huge \begin{array}{l}
\mathbb{R}\\
\mathbb{A}\\
\mathbb{P}\\
\mathbb{P}\\
\mathbb{O}\\
\mathbb{R}\\
\mathbb{T}\\
\\
\mathbb{D}\\
\mathbb{E}\\
\\
\mathbb{R}\\
\mathbb{E}\\
\mathbb{C}\\
\mathbb{H}\\
\mathbb{E}\\
\mathbb{R}\\
\mathbb{C}\\
\mathbb{H}\\
\mathbb{E}\\
\\
\end{array}$
};

 \node [xshift=2cm,yshift=-3cm,scale=8] at (current page.north) {$\mathbb{L\ R\ I}$};
\node [xshift=2cm,thick, rectangle, draw,text width=11.5cm,text centered] at
(current page.center)
{\rule{0pt}{1em}\\ 
  \textbf{\Large{CERTIFIED IMPOSSIBILITY RESULTS FOR BYZANTINE-TOLERANT
    MOBILE ROBOTS}}
\bigskip

{\large AUGER C / BOUZID Z / COURTIEU P / TIXEUIL S /\linebreak  URBAIN X
}
\bigskip

\textbf{Unité Mixte de Recherche 8623\\ CNRS-Université Paris Sud -LRI}
\bigskip

06/2013
\bigskip

\textsf{Rapport de Recherche N$^\circ$ 1560}\\
~

};

 \node [xshift=2cm, yshift=3.5cm,thick, rectangle, draw,text width=11.5cm,text centered] at
 (current page.south)
{\textbf{CNRS - Université de Paris Sud}

Centre d'Orsay

LABORATOIRE DE RECHERCHE EN INFORMATIQUE

Bâtiment 650

91405 ORSAY Cedex (France)};

\end{tikzpicture}

\newpage

\title{Certified Impossibility Results for \\ Byzantine-Tolerant
  Mobile Robots\thanks{This work was supported in part by the Digiteo Île-de-France
project \textsc{Pactole} 2009-38HD.}}

\institute{ 
  {\'Ecole Nat. Sup. d'Informatique pour l'Industrie
    et l'Entreprise (ENSIIE), \'Evry, F-91025}
  \and {\textsc{C\'edric} -- Conservatoire national des arts et
    m\'etiers, Paris, F-75141}
  \and {LRI, CNRS UMR 8623, Universit\'e Paris-Sud, Orsay, F-91405}
% LRI, CNRS UMR 8623, Université Paris Sud & Inria Saclay -- Île-de-France
%  Centre Universitaire d'Orsay, Bâtiment 650 (PCRI), Orsay, F-91405
  \and {UPMC Sorbonne Universit\'{e}s}
\and Institut Universitaire de France
  }
\author{Cédric Auger
\and Zohir Bouzid\inst{4}
\and Pierre Courtieu\inst{2}
\and\\ Sébastien Tixeuil\inst{4,5} 
\and Xavier Urbain\inst{1,3} 
}

\maketitle

\begin{abstract}
  We propose a framework to build formal developments for robot
  networks using the \coq proof assistant, to state and to prove
  formally various properties. We focus in this paper on
  \emph{impossibility} proofs, as it is natural to take advantage of
  the \coq higher order calculus to reason about algorithms as
  abstract objects. We present in particular %the first
  formal proofs of two impossibility results for convergence of
  oblivious mobile robots if respectively more than one half and more
  than one third of the robots exhibit Byzantine failures, starting
  from the original theorems by Bouzid \emph{et al.}. % , thus
  % guaranteeing their soundness.
  Thanks to our formalization, the
  corresponding \coq developments are quite compact. To our knowledge,
  these are the first certified (in the sense of formally proved)
  impossibility results for robot networks.

\end{abstract}

\newpage

\section{Introduction}
\label{sec:introduction}

Networks of static and/or mobile sensors (that is, robots)~\cite{FPS12b} received
increasing attention in the past few years from the Distributed
Computing community. On the one hand, the use of cooperative swarms of
inexpensive robots to achieve various complex tasks in potentially
hasardous environments is a promising option to reduce
human and material costs and assess the relevance of Distributed
Computing in a practical setting. On the other hand, execution model
differences warrant extreme care when revisiting ``classical results''
from Distributed Computing, as very small changes in assumed
hypotheses may completely change the feasibility of a particular problem.
Negative results such as impossibility results are fundamental in
Distributed Computing to establish what can and cannot be computed in
a given setting, or permitting to assess optimality results through
lower bounds for given problems. Two notorious examples are the
impossibility of reaching consensus in an asynchronous setting when a
single process may fail by stopping unexpectedly~\cite{FLP85j}, and the
impossibility of reliably exchanging information when more than one
third of the processes can exhibit arbitrary
behaviour~\cite{PSL80j}. As noted by Lamport~\cite{lamport82jacm}, correctly proving
results in the context of Byzantine (\emph{a.k.a.} arbitrary behaviour
capable) processes is a major challenge, as [they knew] \emph{of no
area in computer science or mathematics in which informal reasoning is more
likely to lead to errors than in the study of this type of
algorithm}.

An attractive way to assess the validity of distributed
algorithm is to use \emph{tool assisted} verification, be it based
process algebra~\cite{bezem97fac,fokkink07}, local
computations~\cite{litovsky99hb}, \texttt{Event-B}~\cite{cansell07},
\coq~\cite{casteran09scss}, HOL~\cite{chou95cj}, Isabelle/HOL~\cite{kuefner12ifiptcs},
or TLA~\cite{lamport82jacm,DBLP:conf/wdag/Lamport11a} that can enjoy
an Isabelle back-end for its
provers~\cite{cousineau12fm}. Surprisingly, only few works consider
using mechanized assistance for networks of mobile entities, be it
population protocols~\cite{deng09tase,clement11icdcs} or mobile robots~\cite{devismes12sss,bonnet12sss}. 
In this paper, our goal is to propose a formal provable
framework in order to prove positive or negative results for
localised distributed protocols in mobile robotic networks,
based on recent advances in mechanical proving and related areas, and
in particular on \emph{proof assistants}. Proof assistants are
environments in which a user can express programs, state theorems and
develop interactively proofs that will be mechanically checked (that
is machine-checked). They have been successfully employed for various
tasks such as the formalisation of programming language
semantics~\cite{Leroy-backend,Mccarthy67correctnessof}, verification
of cryptographic protocols~\cite{DBLP:conf/ccs/AlmeidaBBBKB12},
certification of RSA keys~\cite{thery07tphols}, mathematical
developments as involved as the 4-colours~\cite{Gonthier4color}
or Feit-Thompson~\cite{DBLP:conf/popl/Gonthier13} theorems. 

\subsubsection*{Our contribution}

We developed a general framework relying on the \coq proof assistant
to prove possibility and impossibility results about mobile robotic
networks. 
The key property of our approach is that its underlying calculus is 
of higher order: instead of providing the code of the distributed
protocols executed by the robots, we may quantify universally on 
those programs/algorithms, or
just characterize them with an abstract property.
This genericity makes this approach complementary to the use of
model-checking methods for verifying distributed
algorithms~\cite{cadilhac06avocs,clement11icdcs,devismes12sss} that
are highly automatic, but address mainly particular instances of algorithms.
In particular, quantifying over algorithms allows us to express in a
natural way \emph{impossibility results}.

We illustrate how our framework allows such certification by
providing \coq proofs of two earlier impossibility and lower bound 
theorems by Bouzid \emph{et
  al.}~\cite{bouzid10tcs}, guaranteeing soundness of the first one,
and of the SSYNC fair version of the second one. More precisely, in the
context\footnote{Distributed Robot model assumptions are presented in Section~\ref{sec:robots}.} of oblivious robots that are endowed with strong global
multiplicity detection and whose movements are constrained along a
rational line, and assuming that the demon (that is, the way robots
are scheduled for execution) is fair, the
convergence problem cannot be solved if respectively not less than one
half (Theorem~\ref{thm:demi}) and not less than one third
(Theorem~\ref{thm:tiers}) of robots are Byzantine.

The interestingly short size of the \coq proofs we obtained using our
framework not only makes it easily human-readable, but is also very
encouraging for future applications and extensions of our framework.

\subsubsection*{Related work.}

With reference to proof assistants, K\"ufner \emph{et
al.}~\cite{kuefner12ifiptcs} develop a methodology to develop
\textsc{Isabelle}-checked proofs of properties of fault-tolerant
distributed algorithms in a asynchronous message passing style
setting. This work's motivations are similar to ours, however the
setting (message passing distributed algorithms) is different,
moreover it focuses on {positive} results only whereas we provide
{negative} results, \emph{i.e.} proofs of impossibility.

Chou~\cite{chou95cj} develops a methodology based on the HOL proof
assistant to prove properties of concrete distributed algorithms via
proving simulation with abstract ones. The methodology does not allow
to prove impossibility results.
Casteran \emph{et al.}~\cite{casteran09scss} propose proofs of
negatives results in \coq for some kinds of
distributed algorithms. Though very interesting, their approach is
based on labeled graph rewriting and does not address robot
networks. Another interesting approach is that of Deng and
Monin~\cite{deng09tase} that uses \coq to prove the correctness of
distributed self-stabilizing protocols in the population protocol
model. This model permits to describe interactions of an arbitrary
large size of mobile entities, but the considered entities lack
movement control and geometric awareness that are characteristic of
robot networks such as those we envision, and is thus not suitable for
our purpose. This approach also only considers positive results.

Preliminary attempts for automatically proving impossibility results
in robot networks properties are due to Devismes \emph{et
  al.}~\cite{devismes12sss} and to Bonnet \emph{et
  al.}~\cite{bonnet12sss}. The first paper uses LUSTRE formalism and
model-checking to search exhaustively all possible 3-robots protocols
that explore every node of a $3\times 3$ grid (and conclude that no
such algorithm exists). The second paper uses an ad hoc tool to
generate all possible unambiguous protocols of $k$ robots operating in
an $n$-sized ring ($k$ and $n$ are given as parameters) and
check exhaustively the properties of the generated protocols (and in
the paper conclude that no protocol of $5$ robots on a $10$ sized ring
can explore all nodes infinitely often with every robot). Those two
proposals differ from our goal in several ways. Firstly, they are limited
to a so called \emph{discrete space}, where the robots may only occupy
a \emph{finite} number of positions, while we focus on the more
realistic setting where an infinite number of positions are possible
for the robots. Also, contrary to both, we do not want to restrict our
tools to a particular setting (\emph{e.g.} 3 robots on a $3\times 3$
grid), but rather have results that are general with respect to all
considered parameters. Then, unlike the second proposal, we want
universal impossibility results (\emph{i.e.} consider not only
unambiguous protocols -- that permit to limit combinatorial explosion
to some extend -- but also ambiguous ones -- resulting from
symmetrical situations that are likely to occur in practice). Finally,
we want to integrate the possibility of misbehaving robots
(\emph{e.g.} robots crashing or exhibiting arbitrary and potentially
malicious behaviour), rather than assuming that all considered robots
are correct. This enables to state formally and assess the amount of
faults and attack resilience a given robot protocol may guarantee,
which is crucial when robots are deployed in dangerous areas as it is
often the case.

\subsubsection*{Roadmap.} 
The sequel of the paper is organized as follows.
First, we recall the context of robot networks in Section~\ref{sec:robots}. %
Then, in Section~\ref{sec:proof-assistant} we give a brief
description of \coq and its main principles. %
Section~\ref{sec:formal-model} contains the basis of our formal model
for robot networks, and some useful theorems. %
We show in Section~\ref{sec:some-proofs} how convenient it is to carry
out formal proofs of various properties, as we study previous results by
Bouzid \emph{et al.}~\cite{bouzid10tcs}. %
We provide some concluding remarks in Section~\ref{sec:concl}. %

Note that for the sake of readability we slightly simplified \coq
notations (mostly to avoid syntactic sugar). The actual development
for \coq 8.4pl3 is
available at \url{http://pactole.lri.fr/}

\section{Robot Networks}\label{sec:robots}

We borrow most of the notions in this section
from~\cite{suzuki99siam,agmon2006fault,FPS12b}.
The network consists in a set of $n$ mobile entities, called robots, arbitrarily located 
in the space.
Robots cannot communicate directly by sending messages to each others.
Instead, their communication is based on vision:
they observe the positions of other robots, and based on their observations, 
they compute destination points to which they move.

Robots are \emph{homogeneous} and \emph{anonymous}: 
they run the same algorithm (called \emph{robogram}),
they are completely 
indistinguishable by their appearance, and no 
identifier can be used in their computations.
They are also \emph{oblivious}, {i.e.} they cannot remember any previous observation, computation or movement
performed in any previous step.

For simplicity, we assume that robots are \emph{without volume}, \emph{i.e.} 
they are modeled as points that cannot obstruct the movement or vision of other robots.
Visibility
is \emph{global}: the entire set of robots can always be seen by
any robot at any time.
Robots that are able to determine the exact number of robots occupying a same
position enjoy \emph{strong} multiplicity detection ; if they can
only know if a given position is inhabited or not, their multiplicity
detection is said to be \emph{weak}. 
Each robot has its own local coordinate system and its own unit measure. They do not share
any origin, orientation, and more generally any frame of reference.

The multiset of positions of robots at a given time is called a
\emph{configuration}.
We assume that the actions of robots are controlled by a fictitious entity called the \emph{demon} (or adversary).
Each time a robot is activated by the demon, it executes a complete three-phases cycle: 
Look, Compute and Move.
During the Look phase, using its visual sensors, the robot gets a snapshot of the current configuration. 
Then, based only on this observed configuration, it computes a destination in the Compute phase using its robogram
and moves towards it during the subsequent Move phase.
Movements of robots are \emph{atomic}, \emph{i.e.} the demon cannot stop them before they reach the destination.

A \emph{run} (or execution) is an infinite sequence of rounds. 
During each round, the demon chooses a subset of robots and activates them to execute a cycle.
We assume the scheduling to be \emph{fair}, \emph{i.e.} each robot is
activated infinitely often in any infinite execution, 
and \emph{atomic} in the sense that robots that
are activated at the same round execute their actions synchronously and atomically.
An atomic demon is called fully-synchronous (FSYNC) if all robots are activated at each round, 
otherwise it is said to be semi-synchronous (SSYNC).
The impossibility results we focus on are given in the FSYNC and SSYNC
models, and hence remain valid in less constrained ones (\emph{e.g.}
non-atomic, unfair scheduling, etc.).

A robot is \emph{Byzantine} (or faulty) if it does not comply with the robogram
and behaves in arbitrary and unpredictable way.
We assume that the movements of Byzantine robots are controlled by the adversary 
that uses them in order to make the algorithm fail.
Let $f \in [0,n]$ be a parameter that denotes the number of faulty robots.
Robots that are not Byzantine are called \emph{correct}. Correct
robots are supposed to know an upper bound on the number of Byzantine
robots. 

\section{The \coq Proof Assistant}
\label{sec:proof-assistant}

\coq{} is based on \emph{type theory}.
Its  \emph{formal language} can express objects, properties
and proofs in a unified way; all these are represented as terms of an
expressive $\lambda$-calculus: the \emph{Calculus of Inductive
  Constructions} (CIC)~\cite{coquand90colog}. $\lambda$-abstraction is
denoted \lstinline!fun~x:T~=>~t!, and application is denoted
\lstinline!t~u!.  A proof development with \coq{} consists in trying to
build, interactively and using tactics, a $\lambda$-term the type of
which corresponds to the proven theorem (Curry-Howard style).

The kernel of \coq is a \emph{proof checker} which checks the validity
of proofs written as CIC-terms. Indeed, in this framework, a term is a
\emph{proof} of its type, and checking a proof consists in typing a
term.  Roughly speaking, the small kernel of \coq{} simply
type-checks $\lambda$-terms to ensure soundness.

A very powerful feature of \coq{} is the ability to define
\emph{inductive types} to express inductive data types and inductive
properties. For example the following inductive types define the data
type \lstinline!nat! of natural numbers, \lstinline!O! and
\lstinline!S! (successor) being the two constructors, and the property \texttt{even} of being
an even natural number. In this setting the term %
\texttt{\small even\_S(S(S O))(even\_S O (even\_O))} is of type %
\texttt{\small even(S(S(S(S O))))} so it is a proof that $4$ is even.

\begin{lstlisting}
Inductive nat : Set := O : nat | S : nat -> nat.
Inductive even : nat -> Prop := 
  | even_O : even O 
  | even_S : forall n : nat, even n -> even (S(S n)).
\end{lstlisting}

We also make use of \emph{coinductive} types to express infinite data
types and properties on them. For example in the robot networks setting
a set of robots has an infinite behaviour. For example one can define infinite
streams of natural numbers and the property \texttt{\small all\_even}
of being a infinite stream of even natural number as follows:

\begin{lstlisting}
CoInductive stm : Set :=
  | scons : nat -> stm -> stm.
CoInductive all_even : stm -> Prop :=
  | Ceven_all: forall n s, even n -> all_even s -> all_even (scons n s).
\end{lstlisting}

\section{The formal model}
\label{sec:formal-model}

We present our formal model and the relevant notations. Robots are
anonymous, however we need to identify some of them in the proofs. Thus, we consider the
union of two given disjoint finite sets of \emph{identifiers}: $G$ referring
to robots that behave correctly, and $B$ referring to the set of
Byzantine ones\footnote{We will omit $G$ and $B$ most of the time, except in
  Section~\ref{sec:some-proofs} where they characterise the number of robots.}. Note that those sets are isomorphic to segments of
$\mathbb{N}$ but we keep our formalisation as abstract as possible. If
needed in the model, we can make sure that names are not used by the
embedded algorithm, as shown below.

\begin{lstlisting}
Variable G B : finite.
Inductive ident := Good : G -> ident | Byz : B -> ident.
\end{lstlisting}

\paragraph*{Locations, Positions, Similarities.}
Robots are distributed in space, at places called \emph{locations}.
We define a \emph{position} as a \emph{function} from a set of identifiers to
the space of locations. As the space of locations in the paper of
Bouzid \emph{et al.}~\cite{bouzid10tcs} is an infinite line, we use \setQ
for locations. Note that going from one to many dimensions is not a
problem with respect to our formalisation.  Throughout this article,
and unless specified otherwise $\gp$ denotes a position for
correct robots, and $\bp$ a position for Byzantine ones. The position of
all robots is then given by the combination $\gp \uplus \bp$.

\begin{lstlisting}
Record position:= { gp: G -> location ; bp: B -> location }.
(* Getting the location of a robot *)
Definition locate p (id: ident): location :=
  match id with 
  | Good g => p.(gp) g 
  | Byz b => p.(bp) b end.
\end{lstlisting}

Robots compute their target position from the observed configuration of their
siblings in the considered space. We also define permutations of robots, that is bijective applications from $G\cup B$ to itself, usually denoted hereafter by Greek
letters.
Moreover, any correct robot is supposed to act as any other correct
robot in the same context, that is, with a \emph{similar} perception
of the environment.
For two rational numbers $k \neq 0$ and $t$, a \emph{similarity} is a
function mapping a location $x$ to $k\times (x - t)$, denoted
$\sem{k,t}$. Rational number $k$ is called the homothetic factor, and
$-k\times t$ is called the translation factor. For
simplicity we restrict this definition to the uni-dimensional
case; otherwise rotational factors may have to be provided too.
Similarities are invertible; they form a group for the law of composition
($\sem{k,t}^{-1} = \sem{k^{-1},-k^{-1}\times t}$).
Similarities can be extended to positions, by applying the similarity
transform to the extracted location.
\begin{lstlisting}
Definition similarity (k t : Qc) (p:position) : position := {
  gp := fun n => k * (p.(gp) n - t) ; 
  bp := fun n => k * (p.(bp) n - t) }.
\end{lstlisting}
This operation will be (abusively) written $\sem{k,t} (\gp \uplus
\bp)$. Similarities will be used as transformations of frames of
reference.

\paragraph*{Robograms.}
We now model what an algorithm $r$ embedded in a correct
robot is.
\newcommand{\robid}{\ensuremath{\textsl{r-id}\x}}
For a robot $\robid_i$, a computation takes as an input an entire
position $\gp \uplus \bp$ as seen by $\robid_i$, in its own frame of
reference (scale, origin, etc.),\footnote{Note that the scale factor is taken 
  anew at each cycle for \emph{oblivious} robots; in the context of Byzantine failures, it is
  convenient to consider it as chosen by some adversary.\label{foot:adversary}} and returns a rational number $l_i$
corresponding to a location (the \emph{destination point}) in the same frame.

\begin{remark}\label{rem:automorph}
  Recall that robots in $G$ cannot decide whether another robot is
  Byzantine, and have no access to a symmetry breaking mechanism such
  as an identifier. In such a case: the result of $r$ must be invariant by
  permutations of robots. This is a fundamental property that
  \emph{any} embedded algorithm must fulfil.
\end{remark}

Embedded computation algorithms verifying Remark~\ref{rem:automorph}
are called \emph{robograms}, they are naturally defined in our \coq
model as follows, two sets (i.e. objects of type
\lstinline!finite!).
Note that this definition is completely abstract and makes no use of
concrete code whatsoever.
\begin{lstlisting}
Record robogram := {
 algo : position -> location ;
 AlgoMorph : forall p q sigma, (q == p \o sigma^-1) -> algo p = algo q }.
\end{lstlisting}

\paragraph*{Computation.}\label{sec:formal-computation}

So as to provide to $r$ the locations of robots in terms of the
considered robot's local frame of reference, and to obtain an absolute
location in the \emph{global} coordinate system from
the result of $r$ (thus local) we use the notion of similarity.  Let
us consider a robot $\robid_i$ the location of which is at $t$, and
the scale of which is $k$ times the global one, defining a
similarity $\sem{k,t}$. To obtain the resulting location in terms of
the global coordinate system:
\begin{enumerate}
\item We center the origin of the position in $t$, and
% \item
  we zoom according to the homothetic factor $k$ to express the
  position in 
  the local frame of  $\robid_i$.
\item The algorithm $r$ computes a local destination point.
\item We apply the inverse of the similarity to obtain the global destination
  point, that is: according to the global coordinate system.
\end{enumerate}
We denote this operation $r_{\sem{k,t}}(\gp \uplus \bp) =
\sem{k,t}^{-1}(r(\sem{k,t}(\gp \uplus \bp)))$.
This way we ensure that the global destination point does not depend
on the individual frame of reference of robots.\footnote{Note that in this
  presentation, any considered robot perceives itself as the origin of
  its local frame of reference}

\paragraph*{Demons and Properties.}\label{sec:formal-demons}
  A demon provides the position for Byzantine robots, and selects the
  correct robots to be activated at the current round. As noticed in
  Footnote~\ref{foot:adversary}, we may consider that the demon,
  acting as an adversary, selects also the scale of the frame of
  reference for each activated
  correct robot at each round. A demonic action is thus a record 
\begin{lstlisting}
Record demonic_action:= {locate_byz: B -> location; frame: G -> Qc}.
\end{lstlisting}
consisting of a position for Byzantine robots
(\lstinline!locate_byz!)% 
, and a function associating to each correct
robot a rational number $k$ such that $k = 0$ and the robot is not
activated, or $k \neq 0$ and the robot is activated with a scale
factor.% 
The actual \emph{demon} is simply an infinite sequence (stream) of
demonic actions.
\begin{lstlisting}
CoInductive demon :=  NextDemon: demonic_action -> demon -> demon.
\end{lstlisting}

Characteristic properties of demons include \emph{fairness} and
synchronous aspects.
A demon (seen as a sequence) is locally fair for a robot (inductive property
\lstinline!LocallyFairForOne!) if either this robot is activated
during the first demonic action, or if the robot is not activated
during the first round but the sequel of the demon is locally fair for
that robot. This is related to the classical notion of accessibility.
The demon will be fair if it is locally fair for all robots and if its
\emph{infinite} sequel is fair. 

\begin{lstlisting}
Inductive LocallyFairForOne g (d : demon) : Prop :=
  | ImmediatelyFair : ((demon_head d).frame g) <> 0 
    -> LocallyFairForOne g d
  | LaterFair : ((demon_head d).frame g) = 0  
    -> LocallyFairForOne g (demon_tail d)
    -> LocallyFairForOne g d.

CoInductive Fair (d : demon) : Prop :=
  AlwaysFair : Fair (demon_tail d) 
               -> (forall g, LocallyFairForOne g d) 
               -> Fair d.
\end{lstlisting}

To be fully synchronous\label{page:formal_fsync} for a demon can be defined similarly.  Recall
that a fully synchronous demon is a particular case of fair demon such
that all correct robots are activated at each round. This is done
easily in our setting where we only have to state that the demonic
action's \lstinline!frame! never returns $0$.
An inductive property \lstinline!FullySynchronousForOne! states that
the first demonic action activates a given robot.
A demon is then fully synchronous if
\lstinline!FullySynchronousForOne! holds for all robots and this
demon, and if its \emph{infinite} sequel is fully
synchronous.
\begin{lstlisting}
CoInductive FullySynchronous d :=
  NextfullySynch: FullySynchronous (demon_tail d) 
    -> (forall g, FullySynchronousForOne g d) -> FullySynchronous d.
\end{lstlisting}

\paragraph*{Execution.}\label{sec:execution}
Finally, given an initial position for correct robots $\gp_0$, and a demon
      $$D = (\mbox{\lstinline!locate_byz!}_i, \mbox{\lstinline!frame!}_i)_{i\in \mathbb{N}}$$,
      we may define an infinite sequence $(\gp_i)_{i\in\mathbb{N}}$ called the \emph{execution} (from $\gp_0$
      according to $D$)
      as
      \[\gp_{i+1}(x)=\left\lbrace\begin{array}{ll}
       r_{\sem{\mbox{\scriptsize\texttt{frame}}_i(x),gp_i(x)}}(\gp_i \uplus
       \bp_i) & \textrm{if } \mbox{\texttt{frame}}_i(x) \neq 0 \\
       \gp_i(x) & \textrm{otherwise}\end{array}\right.\]
Its type is thus: 
\begin{lstlisting}
CoInductive execution := 
  NextExecution : (G -> location) -> execution -> execution.
\end{lstlisting}
and its computation is reflected by the following corecursive function
\lstinline!execute!:
\begin{lstlisting}
Definition round 
  (r : robogram) (da : demonic_action) (gp: G -> location) : 
    G -> location := 
  fun g =>
   let k := da.(frame) g in let t := g.(gp) in
   if k = 0 then t 
   else t + (*@$\frac{1}{k}$@*) * (algo r ([[k,t]]{gp := gp; bp := locate_byz da})).

\end{lstlisting}
\begin{lstlisting}
Definition execute (r : robogram): 
  demon -> (G -> location) -> execution :=
    cofix execute d gp :=
      NextExecution gp (execute (demon_tail d) (round r (demon_head d) gp)).
\end{lstlisting}

\section{Case Study: Impossibility Proofs with Byzantine Behaviours}
\label{sec:some-proofs}

Let us illustrate how well-suited our formalisation
is to prove impossibility results, with %by formalising and certifying two
two theorems by Bouzid \emph{et al.}~\cite{bouzid10tcs}.
Those results address the problem known as \emph{convergence}.
Given any initial configuration of robots, the convergence problem
requires \emph{correct} robots to approach asymptotically the
same, but unknown beforehand, location.
That is, for every initial configuration,
convergence requires the existence a point $c$ in space %\marginote{was
                                %$\in \mathbb{R}$} 
such that for every $\epsilon > 0$, there exists a time
$\tau_\epsilon$ such that $\forall \tau > \tau_\epsilon$, all correct
robots are within a distance of at most $\epsilon$ of $c$ at $\tau$.
The impossibility results in~\cite{bouzid10tcs} are as follows:
\begin{theorem}[\cite{bouzid10tcs}, Thm 4.3]\label{thm:demi} It is
  impossible to achieve convergence if $n \leq 2f$ in the
  FSYNC uni-dimensional model,
  where $n$ denotes the number of robots and $f$ denotes the number of
  Byzantine robots.
\end{theorem}

\begin{theorem}[\cite{bouzid10tcs}, Thm 4.4]\label{thm:tiers}
  Byzantine-resilient convergence is impossible for  $n \leq 3f$ in the
  SSYNC
  uni-dimensional model and a \mbox{2-bounded}
  demon.
\end{theorem}

\subsubsection*{Proofs of Impossibility.}\label{sec:impossibility}

Providing a solution to a problem in robot networks usually implies giving
a robogram such that the expected property holds at some point in
the execution, whatever the demon (seen as an adversary, thus including
the Byzantine robots) might do. More precisely, it amounts to showing
that there exists a robogram such that for all demons, the property is
eventually satisfied. An immediate way of proving such a
fact is to provide the actual code for the robogram.

When it comes to impossibility proofs, one has to show instead that
for all robogram pretending to be a solution, there exists a demon
such that the considered robogram will fail. In fact, the usual
attempts to achieve this involve looking for a stronger result:
exhibiting a demon that will make any candidate robogram for solution
to fail. In both cases the statement of such a result is quantified
universally on robograms. Giving any concrete code will not
help. However, working with higher-order mechanical theorem proving
allows to consider programs as abstract objects and to quantify over
them. Robograms will be just characterised by some invariants and the
fact that they are supposed to be a solution of a considered problem.

\subsubsection*{The Theorems in our Formal Model.}\label{sec:formal_theorems}

First of all we need to define formally the convergence problem.
%\paragraph*{Convergence.}\label{sec:convergence}
In the atomic FSYNC and SSYNC models,
an execution $(\gp_i)_{i\in\mathbb{N}}$ is said to be convergent when
for any $\epsilon > 0$ there exists a number of
rounds
$N_\epsilon\in\mathbb{N}$ and a location $l_\epsilon$ (in the
particular context of~\cite{bouzid10tcs}, $l_\epsilon\in\setQ$)
such that for all $n>N_\epsilon$, all correct robots at round $n$
are no further than $\epsilon$ from $l_\epsilon$.
\[\forall \epsilon >0, \exists N_\epsilon\in\mathbb{N}, l\in\setQ,
\forall n>N_\epsilon, \forall x\in G, |\gp_n(x)-l_\epsilon|<\epsilon\]
Convergence expresses that all correct robots will eventually be
gathered forever in a disc of radius $\epsilon$.
That is: robots stay
gathered \emph{forever} in a disc of radius $\epsilon$ (the
coinductive part)\ldots%
\begin{lstlisting}
CoInductive imprisonned (prison_center : location) (radius : Qc)
                        (e : execution) : Prop :=
  InDisk : (forall g, [(prison_center - execution_head e g)] <= radius)
            -> imprisonned prison_center radius (execution_tail e)
            -> imprisonned prison_center radius e.
\end{lstlisting}
\ldots disc that they reach eventually (the inductive part)
\begin{lstlisting}
Inductive attracted (pc: location) (radius: Qc) (e: execution): Prop :=
  | Captured : imprisonned pc radius e -> attracted pc radius e
  | WillBeCaptured : attracted pc radius (execution_tail e) 
                     -> attracted pc radius e.
\end{lstlisting}
A \emph{solution} to the Convergence problem is %thus
a robogram such that for any initial position and assuming a fair demon, the
execution eventually imprisons all correct robots.
\begin{lstlisting}
Definition solution (r: robogram) : Prop := 
  forall (gp: G -> location), forall d: demon, Fair d
  -> forall epsilon: Qc, 0 < epsilon -> exists lim: location, attracted lim epsilon (execute r d gp).
\end{lstlisting}

\begin{remark}\label{rem:limitR}
  Our current model considers locations in \setQ, however the final
  destination (limit) for convergence % needs not to be in $\setQ$, it
  is allowed to be in $\mathbb{R} \setminus \setQ$, in which case the
  sequence of $l_{\epsilon_i}$ is a sequence in $\setQ$ which has a
  limit in $\mathbb{R}$.
\end{remark}

\paragraph{A formal version of Theorem~\ref{thm:demi}.}\label{sec:formal_demi}
Let us focus on Theorem~\ref{thm:demi}. As the premises require the
demon to be fully-synchronous (FSYNC model) we may as well define what
a fully-synchronous demon is, as mentioned on page~\pageref{page:formal_fsync}, and specialise with it a version of
\lstinline!solution!. %
It is worth noticing that our development contains a proof that a
fully-synchronous demon is fair and that therefore a solution for any
fair scheduler is also a solution for a FSYNC one.

\begin{lstlisting}
Definition solution_FSYNC (r : robogram) : Prop := 
  forall (gp : G -> location), forall (d : demon), FullySynchronous d 
  -> forall epsilon: Qc, 0 < epsilon -> exists lim: location, attracted lim epsilon (execute r d gp).
Lemma solution_FAIR_FSYNC : forall r, solution r -> solution_FSYNC r.
Theorem th(*@\ref{thm:demi}@*):
  forall (g b:finite) (g <> \emptyset) -> (r: robogram (singl uplus g) (b uplus (g uplus singl))), 
  \not solution_FSYNC r.
\end{lstlisting}
%Where $g$ is not empty. 
It may seem surprising that we use \texttt{g} both for correct and
Byzantine robots. As a matter of fact, since unions are disjoint by
construction, this notation just ensures that the sets of names share
the same cardinal. Adding another arbitrary set \texttt{b} to the
Byzantine part is thus a way of saying that there are at least as many
Byzantine robots as correct ones.

Further note that this expression of the theorem clearly
states that \emph{there are at least 2 correct robots}; this is not
implicit (as no assumption can be in \coq): the considered set of
correct robots is indeed a singleton added to a
non-empty set.

This theorem and its complete formal proof can be found in our
development, as Theorem \lstinline!no_solution! in File
\lstinline!NoSolutionFSYNC_2f.v!. The file itself is a hundred lines long
and relies on various lemmas provided by our
framework.

\paragraph{A formal SSYNC fair version of Theorem~\ref{thm:tiers}.}\label{sec:formal_tiers}

Akin to the previous
theorem the addition of an arbitrary
set \texttt{b} denotes that the total number of robots is not more
than three times the number of Byzantine ones.

We prove in fact a sligthly different result, instead of assuming
the demon 2-bounded (that is, the demon may execute a particular robot at
most two times between any two executions of \emph{any} other
robot~\cite{DT11r}), we show that the impossibility result holds for a
demon that is fair in SSYNC,
and for a number $f$ of
Byzantine robots such that $2f < n \leq 3f$ where $n$ is the total
number of robots. The bound about $f$ and $n$ by Bouzid \emph{et al.} 
can be obtained by
combining this theorem with the previous one and using lemma
\texttt{\small solution\_FAIR\_FSYNC} above.%

\begin{lstlisting}
Theorem th(*@\ref{thm:tiers}@*)': 
  forall (g b: finite) (g <> \emptyset) -> (r : robogram ((b uplus g) uplus g ) (b uplus g)), 
  \not solution r.
\end{lstlisting}
%\hrule

As before, the theorem and its complete formal proof can be found in our
development, as Theorem \lstinline!no_solution! in File
\lstinline!NoSolutionFAIR_3f.v!. The file itself is 125 lines long
and relies on various lemmas provided by our
framework.

\section{Remarks and Perspectives}
\label{sec:concl}

The choice of the usual topology of $\setQ$ as the basic one is driven
by three main reasons. First, it allows arbitrary homotheties (which is not
the case for $\mathbb{N}$). Then, it preserves arbitrary precision
(thus excluding \textsc{IEEE754} floating point numbers). Finally, it
is axiom-free, while $\mathbb{R}$ is not. As noticed in
Remark~\ref{rem:limitR}, considering rational numbers is not a handicap
for convergence properties.

The total size of our development, including the framework and the
proofs of the aforementioned theorems is quite small, as it is
approximately 450 lines of specifications and 950 lines of proofs. This
is encouraging with reference to how adequate our framework is, as it
indicates that proofs are not too intricate and remain human readable.

It is worth noticing that our formalism is robust enough to take into
account several alternative models with few modifications.  For instance, and
thanks to the high abstraction level of our framework, considering a
multi-dimensional space (instead of just a line) only amounts to
considering tuples for locations (and not simply rational numbers) and
adding a rotation for some similarities. The effort is
thus put on the actual proof and not on the modeling tasks. Hence, a
first short-term perspective is to tackle impossibility proofs for convergence on
the rational plane or three dimensional space. Similarly, going from strong multiplicity to weak
multiplicity is only a redefinition of the equality relation between
positions\ldots The same remark applies to demons'
characteristics. Adding constraints such as being fully-synchronous is
just \emph{(i)} Defining this constraint, and \emph{(ii)} Adding this constraint as an
assumption in the statement of a theorem. Of course proofs may be very
demanding in all those models, but we want to emphasise that relevant
adaptations of our framework are rather non-expensive.

An noteworthy added benefit of our abstract formalisations is
that keeping them as general as possible may lead to relaxing premises
of theorems, thus potentially discovering new results (\emph{e.g.}
formalizing weaker daemons~\cite{DT11r} and weaker forms of Byzantine behaviours
could lead to stronger impossibility results).

Finally, we plan to use our development for positive results also,
that is, to prove properties of concrete algorithms. The language of
\coq can handle data-types, programs, and properties about them. Our
general framework should allow for certification of embedded
algorithms, as both concrete code for robots and global properties of
the network fit in. Notice that such proofs would
guarantee the expected properties in infinite spaces, \emph{i.e.} without
limits on locations.

\newpage

\bibliographystyle{plain}
\bibliography{biblio}

\end{document}